\documentclass[sigconf, nonacm]{acmart}
\usepackage{multirow}
\usepackage[table,xcdraw]{xcolor}
\AtBeginDocument{%
  }

\setcopyright{acmlicensed}
\copyrightyear{2026}
\acmYear{2026}
\setcopyright{cc}
\setcctype{by}
\acmConference[ICCAD '26]{IEEE/ACM International Conference on Computer-Aided Design}{November 08--12, 2026}{San Jose, CA, USA}
\acmBooktitle{IEEE/ACM International Conference on Computer-Aided Design (ICCAD '26), November 08--12, 2026, San Jose, CA, USA}
\acmDOI{10.1145/3831252.3845843}
\acmISBN{979-8-4007-2873-0/2026/11}

\begin{document}

\title{Invited Paper: ShatterQuant: Breaking Uniform Precision\\ with Block-Wise Mixed-Precision on a Systolic Transformer Hardware Accelerator}

\author{Mikolaj Walczak}
\authornote{Both authors contributed equally to this research.}
\email{mwalcza1@jh.edu}
\orcid{0009-0005-0137-0243}
\affiliation{%
  \institution{Johns Hopkins University}
  \city{Baltimore}
  \state{Maryland}
  \country{USA}
}

\author{Edward Humes}
\email{ehumes2@jhu.edu}
\orcid{0009-0002-3945-0116}
\authornotemark[1]
\affiliation{%
  \institution{Johns Hopkins University}
  \city{Baltimore}
  \state{Maryland}
  \country{USA}
}

\author{Chao Fang}
\email{chao.fang@esat.kuleuven.be}
\orcid{0000-0003-3430-1189}
\affiliation{%
  \institution{KU Leuven}
  \city{Leuven}
  \country{Belgium}
}

\author{Marian Verhelst}
\email{marian.verhelst@kuleuven.be}
\orcid{0000-0003-3495-9263}
\affiliation{%
  \institution{KU Leuven}
  \city{Leuven}
  \country{Belgium}
}

\author{Tinoosh Mohsenin}
\email{tinoosh@jhu.edu}
\orcid{0000-0001-5551-2124}
\correspondingauthor
\affiliation{%
  \institution{Johns Hopkins University}
  \city{Baltimore}
  \state{Maryland}
  \country{USA}
}

\renewcommand{\shortauthors}{Walczak et. al.}
\newcommand{\sys}{ShatterQuant}

\begin{abstract}
Due to limited support for intra-tensor heterogeneous precision in conventional accelerators, neural network quantization remains largely restricted to per-tensor precision assignment. We present \sys{}, a hardware-software co-designed framework enabling mixed-precision quantization within each tensor by assigning independent bit-widths to blocks of a weight projection. \sys{} couples precision granularity with PE configuration, such that each precision determines an effective block height. We introduce (1) a hardware-aware post-training method that assigns intra-tensor precision based on block-level standard deviation and weight sensitivity; (2) the \sys{} Transformer Accelerator supporting 1/2/4/8-bit weight precision, precision-dependent PE configuration, block rescaling, and integrated softmax and piecewise-linear nonlinearities; and (3) an evaluation of model-hardware tradeoffs using an implementation in the TSMC 16nm PDK operating at 1 GHz, achieving 1.5 TOPS, 760 GOPS/$mm^2$ area efficiency, and 2.8 TOPS/W energy efficiency. On DeiT and ImageNet-1K, \sys{} achieves accuracy within $3.3\%$ of state-of-the-art mixed-precision techniques while using a 2 bit lower effective bitwidth, while for PixelDiT demonstrates comparable generation quality. \sys{} demonstrates how fine-grained intra-tensor mixed-precision can be realized through hardware-software co-design.
\end{abstract}
% Then, we formulate precision assignment and quantization as a block-reconstruction problem, for which we propose a solution which integrates into existing models via post-training mixed-precision quantization.



\begin{teaserfigure}
 \centering\includegraphics[width=\textwidth]{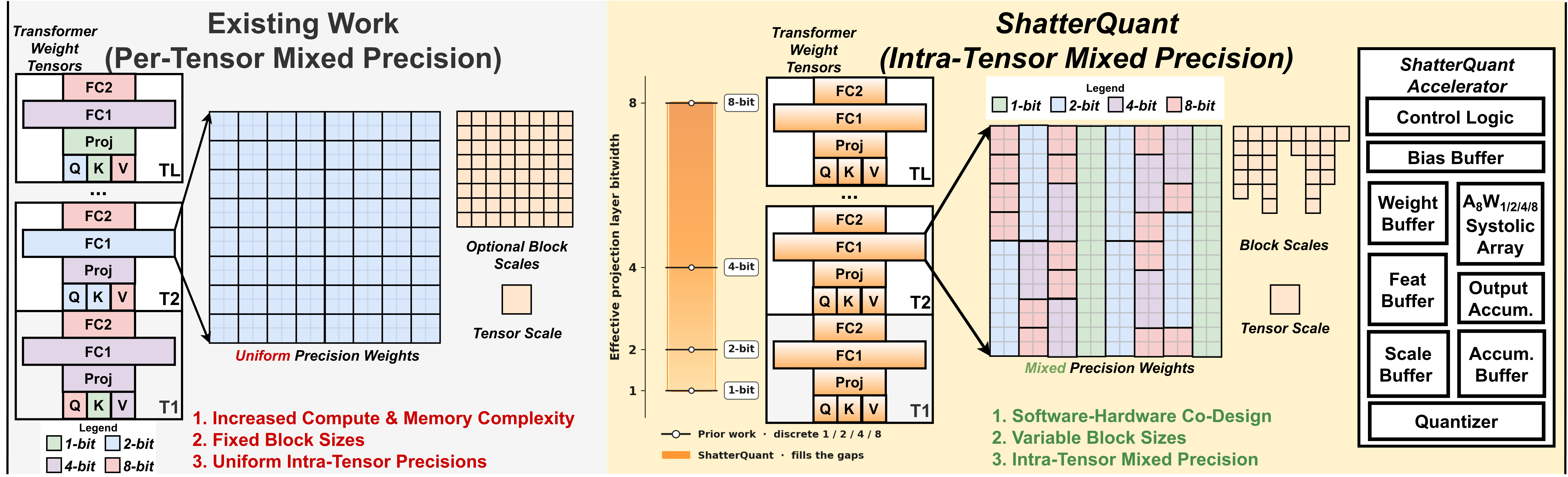}
  \caption{Motivation of \sys{} compared with existing mixed-precision quantization approaches. Existing methods~\cite{Ranjan_2026_CVPR, SUN2026132856} evaluate coarse-grained, tensor-wise precision assignments, whereas \sys{} enables fine-grained mixed-precision assignments within individual weight tensors through hardware-software co-design.}
  % \Description{Enjoying the baseball game from the third-base
  % seats. Ichiro Suzuki preparing to bat.}
  \label{fig:teaser}
\end{teaserfigure}

% \received[accepted]{24 August 2026}

\maketitle

\section{Introduction}
Although transformers~\cite{vaswani2017attention} originated in natural language processing, they have become a foundational architecture across many fields, including computer vision~\cite{liu2021swin,touvron2021training,dosovitskiy2020image} and generative modeling~\cite{10377858,ma2024sit}. Vision Transformers (ViTs) have achieved strong performance in image recognition, while Diffusion Transformers (DiTs) have enabled high-quality image generation. However, these improvements come at the cost of substantial computation and memory traffic. Transformer inference repeatedly executes large matrix and attention operations, with diffusion models further amplifying costs by evaluating the transformer backbone across multiple denoising steps~\cite{10.1145/3586074}. Recent efforts to reduce the computational and memory requirements of transformer models have explored a variety of strategies~\cite{10.1145/3829079}, including architecture search~\cite{edward-dac-2026}, knowledge distillation, pruning~\cite{uttej-2025-TECS-Magrip}, and state-space architectures~\cite{romina-2026-tbiocas-MedMambaLite-v2, ICLR2026_8abd2043, hatamizadeh2026gateddeltanet2decouplingerase}.
%While these techniques reduce model complexity and improve deployment efficiency, inference remains dominated by high-dimensional matrix operations and parameter movement, with latency governed by memory access rather than computation.

Quantization is a promising approach to reduce transformer deployment costs by representing weights and activations using low numerical precision~\cite{ ICLR2026_5c802ef3, cheng2026signroundv2closingperformancegap, JMLR:v26:24-2050}. Uniformly applying low-precision quantization therefore significantly reduces model size, data movement costs, and arithmetic complexity~\cite{11527277}. However, aggressive quantization, particularly at sub-8-bit widths, often causes unacceptable degradation in classification accuracy or generation quality. In the case of transformers, this is caused by parameters exhibiting nonuniform sensitivity to quantization: different layers, channels, and parameter groups may require different numerical precision. Block-wise quantization addresses this limitation by partitioning tensors into smaller groups that can be independently scaled~\cite{dettmers20228bitoptimizersblockwisequantization}. Compared with tensor-wise quantization, this finer granularity provides greater representational flexibility, allowing each block to use a scale better suited to its numerical characteristics while enabling more aggressive overall precision reduction.

Despite the advantages of block-wise quantization in retaining accuracy at reduced precision, current inference platforms such as GPUs often struggle to efficiently support mixed-precision blocks due to non-uniform memory accesses and fixed-function compute datapaths. Therefore, in this work, we present \sys{} a hardware–software codesign framework for mixed-precision block-wise quantization and inference of transformer models. The framework combines a hardware-aware post-training quantization (PTQ) approach with a custom \sys{} Transformer Accelerator implemented in TSMC's 16nm PDK, enabling precision assignments tailored to the accelerator's hardware characteristics. Our key contributions are summarized as follows:
\begin{itemize}

\item We develop a post-training quantization method for Vision Transformers and Diffusion Transformers that assigns mixed-precision bit widths at the block level, enabling aggressive model compression while preserving for both image classification accuracy and generation quality and evaluate the approach on DeiT ViTs and PixelDiT DiT using Imagenet-1k.

\item We design the \sys{} Transformer Accelerator, which supports 1-, 2-, 4-, and 8-bit integer weights with 8-bit activations. The accelerator couples quantization block size with processing-element (PE) configuration, allowing precision to directly determine effective block dimensions and compute parallelism for joint hardware-software configuration.

\item We implement the \sys{} Transformer Accelerator in TSMC's 16nm process design kit, directly mapping quantization block sizes to their corresponding PE configurations. We evaluate the resulting design trade-offs in model quality, hardware area, throughput, energy efficiency, and deployment cost across DeiT image classification and PixelDiT image generation on ImageNet-1k.

\end{itemize}
\section{Related Work}
\subsection{Diffusion Transformers}
DiTs replace the convolutional U-Net backbone traditionally used in diffusion models with a transformer architecture, demonstrating that scaling depth, width, and token count leads to improved generative performance while maintaining a simple design~\cite{10377858}. Building on this foundation, SiT introduces a scaling strategy that jointly optimizes diffusion training efficiency and sample quality through improved transformer design and training methodologies~\cite{ma2024sit}. More recently, PixelDiT further improves transformer-based diffusion by operating directly in the pixel domain, eliminating the need for a separate latent autoencoder while maintaining generation quality and scalability~\cite{yu2026pixeldit}. With the removal of the latent-space autoencoder, we design the \sys{} transformer accelerator to focus on the core components of traditional ViT~\cite{dosovitskiy2020image} architectures used in DeiT~\cite{touvron2021training}. Specifically, \sys{} targets linear projections, attention softmax, and nonlinear activation functions.

\subsection{Mixed Precision Quantization}
Transformer quantization has emerged as a key technique for reducing the computational and memory costs of large-scale models. GPTQ~\cite{frantar2023gptqaccurateposttrainingquantization} introduced an effective PTQ framework for transformer weights, enabling low-bit deployment while maintaining high accuracy. Subsequent work has explored quantization-aware training and mixed-precision strategies to better preserve model performance for low-bit vision transformers~\cite{walczak2025bitmedvit, JMLR:v26:24-2050}, improved PTQ through randomized quantization during calibration~\cite{wei2023qdroprandomlydroppingquantization}, and recent mixed-precision techniques that adapt bitwidths according to layer sensitivity~\cite{Hwang_2026_CVPR, Ranjan_2026_CVPR, SUN2026132856} or explore channel-wise intra-tensor quantization~\cite{10973048, feng2025mpq}. In contrast, \sys{} enables grouped sub-channel quantization with dedicated hardware support. %Earlier studies also demonstrated the effectiveness of data-free PTQ laying the foundation for many modern quantization approaches~\cite{pmlr-v119-nagel20a}. 

\begin{figure*}[t]
    \centering
    \includegraphics[width=0.85\linewidth]{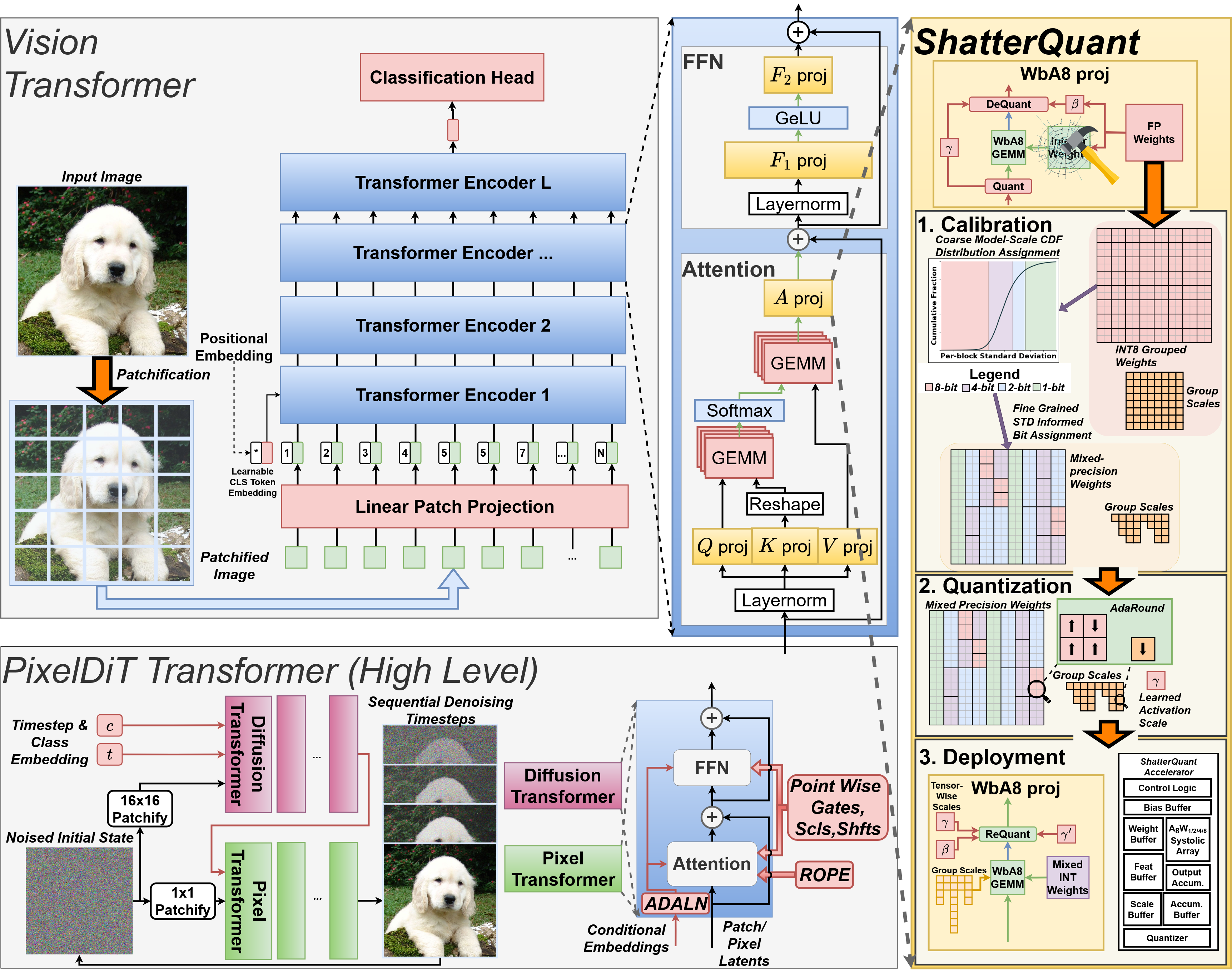}
    \vspace{-2ex}
    \caption{Vision Transformer (ViT) architecture and high-level overview of \sys{}, with PixelDiT shown as a representative vision diffusion model. \sys{} is applied to linear projections through a three-stage process: (1) calibration and initial weight setting, (2) quantization and weight rounding correction, and (3) hardware deployment.
}
    \label{fig:overview}
\end{figure*}

\subsection{Hardware Accelerators}

Specialized hardware accelerators have become increasingly important for executing transformer workloads by reducing memory movement and accelerating compute-intensive operations. Prior work has explored accelerator designs tailored for low-bit transformer inference~\cite{10.1145/3676536.3676681, lin2025systolicarraybasedarchitecturelowbit}, including support for hardware-efficient power-of-two quantization and lightweight arithmetic to improve throughput and energy efficiency~\cite{shi2024p2vitpoweroftwoposttrainingquantization, 11218199}. More recent architectures further optimize transformer execution through specialized mixed-precision dataflows for sparsity~\cite{11148266, 11434978}, on-chip memory hierarchies, and dedicated support for attention mechanisms and mixed-precision computation~\cite{10946730,11261796, 10547183, 11059915}. These works demonstrate the benefits of hardware-software co-design, but largely rely on coarse-grained precision configurations, exploit sparsity by skipping operations rather than varying the weight precision, or contain limited evaluation on modern transformer architectures.

\section{Proposed Approach}
The proposed approach, illustrated within Fig.~\ref{fig:overview}, is separated into four subsections: Model Architecture, Intra-Tensor Blockwise Mixed Precision, Post Training Blockwise Quantization, and the \sys{} Transformer Accelerator.
\subsection{Model Architecture}

The formulation of \sys{} begins with the original ViT~\cite{dosovitskiy2020image}, which consists of $L$ transformer layers. A $C$-channel input image is partitioned into $n$ patches, represented as $P\in\mathbb{R}^{n\times(p^2 \cdot C)}$, and linearly projected into the embedding dimension $D$ using a learnable projection matrix $W_{ED}\in\mathbb{R}^{(p^2C)\times D}$, concatenated with an optional class token ($CLS$), and have a learned positional embedding added to them, producing the initial token embeddings ($X$).

Each transformer layer projects the embeddings $X$ into queries (Q), keys (K), and values (V) using weights $W_Q,W_K,W_V\in\mathbb{R}^{D\times D}$:
\begin{equation}
Q=XW_Q,\qquad K=XW_K,\qquad V=XW_V.
\end{equation}
The projections are divided into $H$ attention heads of dimension $d_k=D/H$. For each head $h$, scaled dot-product attention computes context-aware token representations as
\begin{equation}
\mathrm{Attn}_h=
\operatorname{softmax}\left(\frac{Q_hK_h^\top}{\sqrt{d_k}}\right)V_h,
\end{equation}
The head outputs are concatenated and projected with $W_O\in\mathbb{R}^{D\times D}$:
\begin{equation}
Z=\operatorname{Concat}_{h=1}^{H}(\mathrm{Attn}_h)W_O
\in\mathbb{R}^{N\times D}.
\end{equation}
and processed by a feed-forward network (FFN):
\begin{equation}
\mathrm{FFN}(Z)=\sigma(ZW_1+b_1)W_2+b_2,
\end{equation}
where $W_1\in\mathbb{R}^{D\times D_{ff}}$, $W_2\in\mathbb{R}^{D_{ff}\times D}$, $D_{ff}$ typically set to $4D$, with $\sigma$ as a nonlinear activation function. Together, these components are coupled with residuals and LayerNorm, defining a single transformer layer as illustrated in Fig.~\ref{fig:overview}.

PixelDiT~\cite{yu2026pixeldit} extends the conventional transformer architecture for diffusion-based image denoising by introducing separate patch-wise and pixel-wise transformer pathways. In addition to the standard transformer operations, PixelDiT incorporates adaptive layer normalization (AdaLN), which uses conditioning information to adaptively scale and shift intermediate representations toward the desired class output. Despite these architectural additions, the core transformer computation remains consistent, relying on linear projections and attention operations. We scope our approach applicable to both conventional ViTs and PixelDiT targeting the full attention pipeline and feedforward structure. With this approach we offload operations such as the initial embedding linear and final linear operations, LayerNorm, RoPE, reshaping, and scaling to external processors, while handling the parameterized linear and attention operations on our \sys{} Transformer Accelerator.

\subsection{Intra-Tensor Blockwise Mixed Precision}
To formulate the intratensor blockwise mixed precision weights, we first apply tensor-wise AbsMax scaling to the weights, defining $\beta=\max_{i,j}|W_{ij}|$ as the weight scale with $\tilde{W}=W/\beta$. Activations are quantized to a fixed 8-bit range using a tensor-wise scale $\gamma$, initialized at $\max_{i,j}|X_{ij}|/2^{7}$ and then calibrated to minimize each layer's output error, while the normalized weights $\tilde{W}$ are quantized using block-specific precision and scale. For hardware efficiency, each block scale is constrained to a power of two and represented by a signed 6-bit exponent $k_{r,c}$ and 2-bit precision $b_{r,c}$:
\begin{equation}
s_{r,c}=\frac{2^{k_{r,c}}}{2^{b_{r,c}-1}},
\qquad
q_{\min}=-2^{b_{r,c}-1},
\qquad
q_{\max}=2^{b_{r,c}-1}-1.
\end{equation}
Writing $(r,c)$ for the block containing element $(i,j)$, the weights are quantized as
\begin{equation}
W^{\mathrm{mix}}_{ij}
=
\mathrm{clamp}\!\left(
\mathrm{round}\!\left(
\frac{\tilde{W}_{ij}}{s_{r,c}}
\right),
q_{\min},q_{\max}
\right),
\qquad
W_{ij}
\approx
W^{\mathrm{mix}}_{ij}s_{r,c}\beta .
\end{equation}
As represented in Fig.~\ref{fig:overview}, we make use of non-uniform blocks supporting 8-, 4-, 2-, or 1-bit weights. At 8 bits, each stored weight corresponds to a single output column, while decreasing precision packs two, four, or eight weights, and thus output columns, into the same 8-bit representation. In contrast, activations are quantized to a fixed 8-bit range using the tensor-wise scale $\gamma$:
\begin{equation}
X^{\mathrm{int}}_{ij}
=
\mathrm{clamp}\!\left(
\mathrm{round}\!\left(
\frac{X_{ij}}{\gamma}
\right),
-128,127
\right),
\qquad
X_{ij}
\approx
X^{\mathrm{int}}_{ij}\gamma .
\end{equation}

\subsection{Post-Training Blockwise Quantization}

In order to efficiently quantize pretrained models, we treat each weight block as being part of a ``superblock,'' where one superblock occupies the same number of weights as a single 1-bit block. Since each block contains a local scale, precisions are determined based on the overall standard deviation (STD) across all superblocks within the selected model using cutoffs within the cumulative distribution function (CDF). As illustrated within stage 1 of Fig.~\ref{fig:overview}, based on the CDF, superblocks with greater STD are coarsely represented using fewer bits, while superblocks with lower STD are represented using higher bitwidths in a fine-grained manner.

To improve model performance we adopt the LS-ViT~\cite{Hwang_2026_CVPR} quantization algorithm, which combines Hessian estimation and reconstruction components~\cite{wei2023qdroprandomlydroppingquantization}, and we adapt its usage of AdaRound~\cite{pmlr-v119-nagel20a} weight rounding to align with the \sys{} weight quantization blockwise structure. In addition to AdaRound's per-weight learned rounding decisions, we extend the weight scaling to the power-of-two block scales $k$, as rounding to the nearest power-of-two can result in lost resolution. During layer reconstruction, alongside the per-weight rounding decision variables $u_{ij}$, each superblock contains a single $v_m$ which rounds between $2^k$ and $2^ {k-1}$. Thus, the reconstructed weight is

\begin{equation}
\hat{W}=\hat{W}^{(k)}+h(v_m)\left(\hat{W}^{(k-1)}-\hat{W}^{(k)}\right).
\end{equation}

Both candidates are fully quantized, so the exponent is selected without relaxing the power-of-two constraint itself, and both decisions are annealed by the same regularizer against the least-squares Hessian objective.

\subsubsection{Extension to PixelDiT}
We reuse our LS-ViT adaptation for PixelDiT, however, we make several changes. First, we sample a set of image generation trajectories from the full precision model, drawing from multiple classes during the process. We utilize a squared error between the full precision model's outputs and quantized model's outputs at each timestep as the task loss for LS-ViT's Hessian estimation. Second, as prior works~\cite{li2023q} have noted, the activation distributions of diffusion and flow matching models tend to shift over each inference timestep. Therefore, we further extend the LS-ViT adaptation by initially calibrating multiple activation scales per layer, separated into 8 timestep bins, with the active scale selected by timestep. Third, rather than using STD and a CDF-based cutoff, we instead assign precisions via a relative variance score, normalizing each superblock's variance by the mean variance of its own layer, along with a fixed percent setting a cutoff for the superblocks with greatest relative variance to get high precision, and vice versa. Finally, although we still optimize the per-layer activation scales during reconstruction, we add a post-reconstruction stage which optimizes only the activation scales across all layers to minimize MSE from the FP model trajectories.

\begin{figure*}[t!h]
    \centering
    \includegraphics[width=0.85\linewidth]{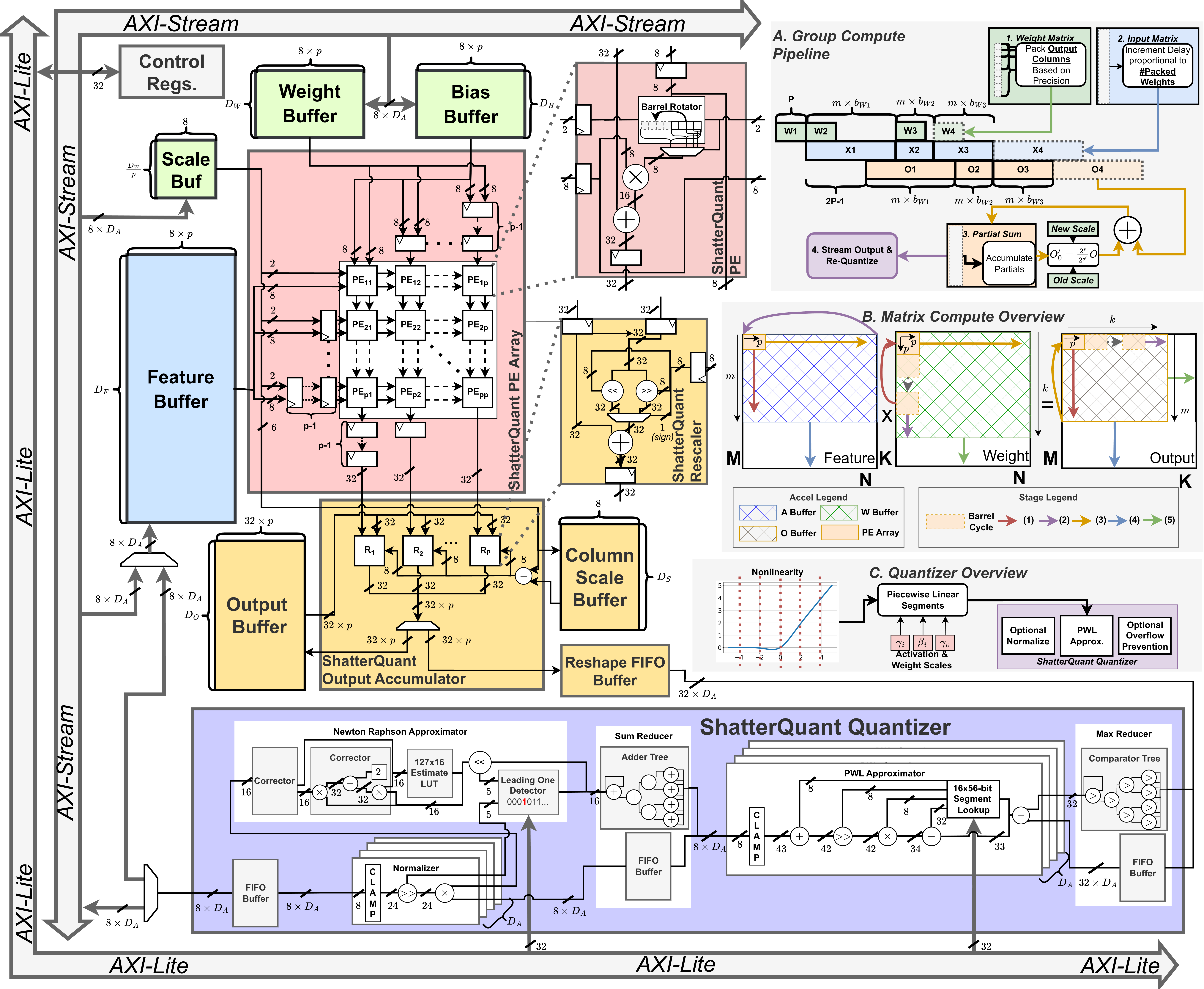}
    \vspace{-2ex}
    \caption{Overview of the \sys{} Transformer Accelerator and dataflow. Green: weight memory and input scales; Blue: activation memory; Red: mixed-precision systolic array with barrel-rotator PEs; Yellow: output buffer and rescaler; Purple: fused PWL-based quantizer with softmax normalization and range-overflow prevention. (A) Single-block execution (B) Execution for larger matrix operations. (C) Fused PWL-softmax quantization integrated within the \sys{} quantizer.}
    \label{fig:hardware_architecture}
\end{figure*}

\subsection{\sys{} Transformer Accelerator}
The \sys{} Transformer Accelerator architecture is designed to execute transformer workloads aligned with the \sys{} quantization method. As shown in Fig.~\ref{fig:hardware_architecture}, the architecture consists of a weight, scale, bias, and feature input buffers, an 8-bit activation and 1/2/4/8-bit systolic processing array, a 32-bit partial product accumulator, and an integrated quantizer. The accelerator is controlled via a standard AXI-Stream bus for data streaming and AXI-Lite bus for configuration and initialization.  We break this architecture down into four subsections: processing element (PE), PE array, output accumulator, and quantizer.

\begin{figure*}
    \centering
    \includegraphics[width=\linewidth]{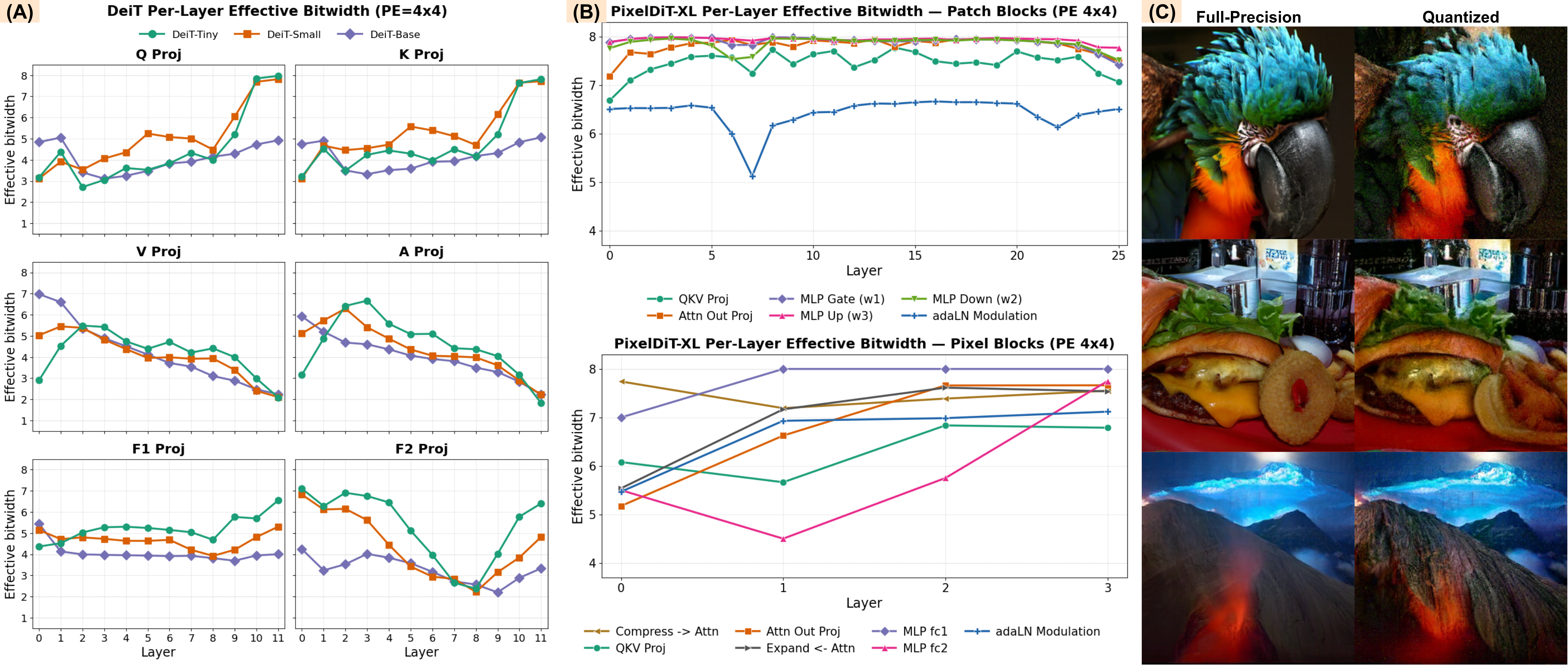}
    \caption{Per layer effective-bitwidth evaluation for DeIT and PixelDiT-XL post \sys{} quantization (A) Effective per-layer bitwidth for DeiT with a 5\% overall 1-bit weight target for a $4\times4$ PE configuration. (B) Effective per-layer bitwidths for PixelDiT-XL for individual projections within the patch \& pixel transformer blocks with 80\% of its weights being 8-bit, and 20\% of its weights being 4-bit, using a $4\times4$ PE array configuration. (C) Comparison of generated images between original full precision PixelDiT-XL model, and quantized}~\vspace{-4ex}
    \label{fig:deit_precision}
\end{figure*}

\subsubsection{\sys{} Processing Element}
The \sys{} PE represents the lowest level weight-stationary multiply-and-accumulate (MAC) engine capable of supporting 1-, 2-, 4-, and 8-bit weight representation. An integrated barrel rotator with a 2-bit rotation select, dynamically extends and rotates lower-bit weight blocks from the packed 8-bit representation, enabling the same PE to operate across different precision modes while using a unified 8-bit multiplier accumulated into a 32-bit result. For 8-bit weights, the PE performs a standard weight-stationary MAC operation. As precision decreases, the PE exploits increased bit-level serialization by reusing input features across multiple clock cycles. Specifically, 4-bit weights maintain the same input feature stream while generating two output columns over two clock cycles, while 2-bit and 1-bit modes expand this to four and eight output columns, respectively.

\subsubsection{\sys{} PE Array}
For the \sys{} PE to support larger matrix sizes at higher throughput, PE's are tiled across a square $p\times p$ systolic array, with weights and biases propagated down the array, features and weight-precision metadata propagated across, and partial sums accumulated at the bottom into the output accumulator. Since each PE includes a dedicated barrel rotator, an additional weight path enables weight preloading without stalling computation, requiring only a $p$ clock cycle weight preload latency and $2p-1$ feature clock cycle initial setup latency. 

To sustain bandwidth, the weight, bias, and feature buffers are organized with $8\times p$-bit-wide interfaces and tunable depths of $D_W$, $D_B$, and $D_F$, respectively. Each scale corresponds to a $p\times p$ block of packed weights; thus, the scale buffer has depth $D_W/p$, with the two lowest bits encoding the block precision and the six highest bits encoding its scale. This organization establishes a direct relationship between the software block size and the systolic core configuration: as weight precision decreases, the effective processing width expands by a factor of two, increasing the number of output columns computed per cycle. %Consequently, \sys{} jointly exposes model precision and array configuration as tunable design parameters, enabling the precision granularity and parallelism to be co-optimized for the desired accuracy and throughput.

\subsubsection{\sys{} Output Accumulator}
The output accumulator maintains 32-bit partial sums and incorporates the \sys{} rescaler for cross-block accumulation. It comprises of a $32\times p$-wide buffer with depth $D_O$ and an 8-bit-wide column scale buffer with tunable depth $D_S$. Rather than dequantizing each block, the rescaling unit adjusts the accumulator scale at block boundaries, enabling integer-domain accumulation. Let $z_i$ denote the integer partial sum from block $i$, with power-of-two scale $s_i$ and corresponding dequantization scale $\Delta_i = 2^{s_i}$ rescaled to the next block's scale using
\begin{align}
\hat{O}_{i+1}
&=
\left(
\hat{O}_i
\frac{\Delta_i}{\Delta_{i+1}}
\right)
+ z_{i+1},
&
\frac{\Delta_i}{\Delta_{i+1}}
&=
2^{s_i-s_{i+1}}.
\end{align}
Thus, rescaling is implemented as a shift by $s_i-s_{i+1}$: a positive value applies a right shift, while a negative value applies a left shift with the scale s$_{i+1}$ stored within the column scale buffer for future block rescaling.

\subsubsection{\sys{} Quantizer}
To enable fully integer inference, the accelerator integrates an on-chip quantizer that performs requantization and nonlinear operations without floating-point arithmetic~\cite{10745859}. This comprises a dedicated max-subtraction tree, piece-wise linear (PWL) approximator, accumulation tree, and Newton--Raphson reciprocal unit~\cite{BURKHARDT1981235}. Because nonlinear inputs originate from the $int32$ accumulator, conventional LUT-based implementations are impractical due to their potentially $2^{32}$-entry address space. Instead, the accelerator implements a compact PWL representation~\cite{11130197}. Specifically, each function is approximated over a set of input intervals as
\begin{align}
f(x) &\approx m_n(x-x_{n-1})+b_n,\quad x_{n-1}\leq x<x_n,
\end{align}
given $m_n$, $b_n$, and $x_n$ denote the segment slope, bias, and boundary. The preceding dequantization scales $\gamma\beta$ and subsequent quantization scale $1/\gamma'$ are fused into the PWL slope, such that
\begin{align}
m_n\frac{\gamma\beta}{\gamma'}
\approx
\frac{s_n}{2^{\kappa_n}},
\end{align}
where $s_n$ and $\kappa_n$ are an integer multiplier and power-of-two shift, respectively. The resulting integer approximation is
\begin{align}
\hat{f}(x)
\approx
\mathrm{RoundClip}\left(
\left(s_n(\hat{x}-\hat{x}_{n-1})\right)\gg\kappa_n+\hat{b}_n
\right).
\end{align}
Each segment requires an $int32$ boundary and $int8$ scale, shift, and bias, which fuse nonlinearities directly into the requantization stage. For softmax, the quantizer implements the approximation~\cite{MLSYS2026_ea5ffdf7}
\begin{align}
\mathrm{softmax}_{INT8}
= \frac{128 e^{x_i-\max(x)}}{\sum_j 128 e^{x_j-\max(x)}} <<s_O,
\end{align}
where $2^{s_O}$ defines the output scale and $128$ scales the bounded exponential to the $int8$ range. The $128$ factor is fused into a PWL approximation of $e^x$, while $s_O$ represents a tunable shift on the reciprocal output and serves as the input scale for following layers. 

\section{Results}
We separate evaluation into software and hardware performance, followed by a study of the relationship between accelerator configuration and software performance. 'PE' is used to describe configurations in both the software and hardware evaluations.

\begin{table}[]
\begin{tabular}{c|cc|cc|cc}
\bottomrule
\rowcolor[HTML]{EFEFEF} 
\cellcolor[HTML]{EFEFEF} & \multicolumn{2}{c|}{\cellcolor[HTML]{EFEFEF}\textbf{DeiT-Tiny}} & \multicolumn{2}{c|}{\cellcolor[HTML]{EFEFEF}\textbf{DeiT-Small}} & \multicolumn{2}{c}{\cellcolor[HTML]{EFEFEF}\textbf{DeiT-Base}} \\ \cline{2-7} 
\rowcolor[HTML]{EFEFEF} 
\multirow{-2}{*}{\cellcolor[HTML]{EFEFEF}\textbf{Method}} & \multicolumn{1}{c|}{\cellcolor[HTML]{EFEFEF}\textbf{Prec.}} & \textbf{Acc.} & \multicolumn{1}{c|}{\cellcolor[HTML]{EFEFEF}\textbf{Prec.}} & \textbf{Acc.} & \multicolumn{1}{c|}{\cellcolor[HTML]{EFEFEF}\textbf{Prec.}} & \textbf{Acc.} \\ \toprule
Baseline & \multicolumn{1}{c|}{FP32} & 74.5 & \multicolumn{1}{c|}{FP32} & 81.2 & \multicolumn{1}{c|}{FP32} & 83.4 \\ \midrule
P2ViT & \multicolumn{1}{c|}{8} & 69.6 & \multicolumn{1}{c|}{8} & 77.6 & \multicolumn{1}{c|}{8} & 80.6 \\
Mix-QViT & \multicolumn{1}{c|}{6} & 71.7 & \multicolumn{1}{c|}{6} & 79.6 & \multicolumn{1}{c|}{6} & 81.6 \\
SLIQ-ViT & \multicolumn{1}{c|}{6} & 71.8 & \multicolumn{1}{c|}{6} & 79.6 & \multicolumn{1}{c|}{6} & 81.8 \\ \midrule
\textbf{Ours} & \multicolumn{1}{c|}{5} & 68.5 & \multicolumn{1}{c|}{4} & 78.2 & \multicolumn{1}{c|}{4} & 80.7 \\ \midrule
\end{tabular}
% \begin{tabular}{c|c|c|c|c}
% \bottomrule
% \rowcolor[HTML]{EFEFEF} 
% \textbf{Method} & \textbf{\begin{tabular}[c]{@{}c@{}}Prec. \\ (A/W)\end{tabular}} & \textbf{T} & \textbf{S} & \textbf{B} \\ \toprule
% Baseline & 32/32 & 74.5 & 81.2 & 83.4 \\ \midrule
% P2ViT~\cite{shi2024p2vitpoweroftwoposttrainingquantization} & 8/MP8 & 69.6 & 77.6 & 80.6 \\
% Mix-QViT~\cite{Ranjan_2026_CVPR} & MP6/MP6 & 71.7 & 79.6 & 81.6 \\
% SLIQ-VIT~\cite{SUN2026132856} & MP6/MP6 & 71.8 & 79.6 & 81.8 \\ \midrule
% Ours & 8/MP(5|4|4) & 68.5 & 78.2 & 80.7\\ \bottomrule
% \end{tabular}
\caption{Comparison of \sys{} against existing mixed-precision works on three DeiT~\cite{touvron2021training} variants shown for Imagenet-1k Top-1 accuracy using the 4x4 PE configuration. Precision defines effective overall model weight bitwidth.}\label{tab:software_comparisons}~\vspace{-7ex}
\end{table}

\begin{figure*}[t]
    \centering
    \includegraphics[width=\linewidth]{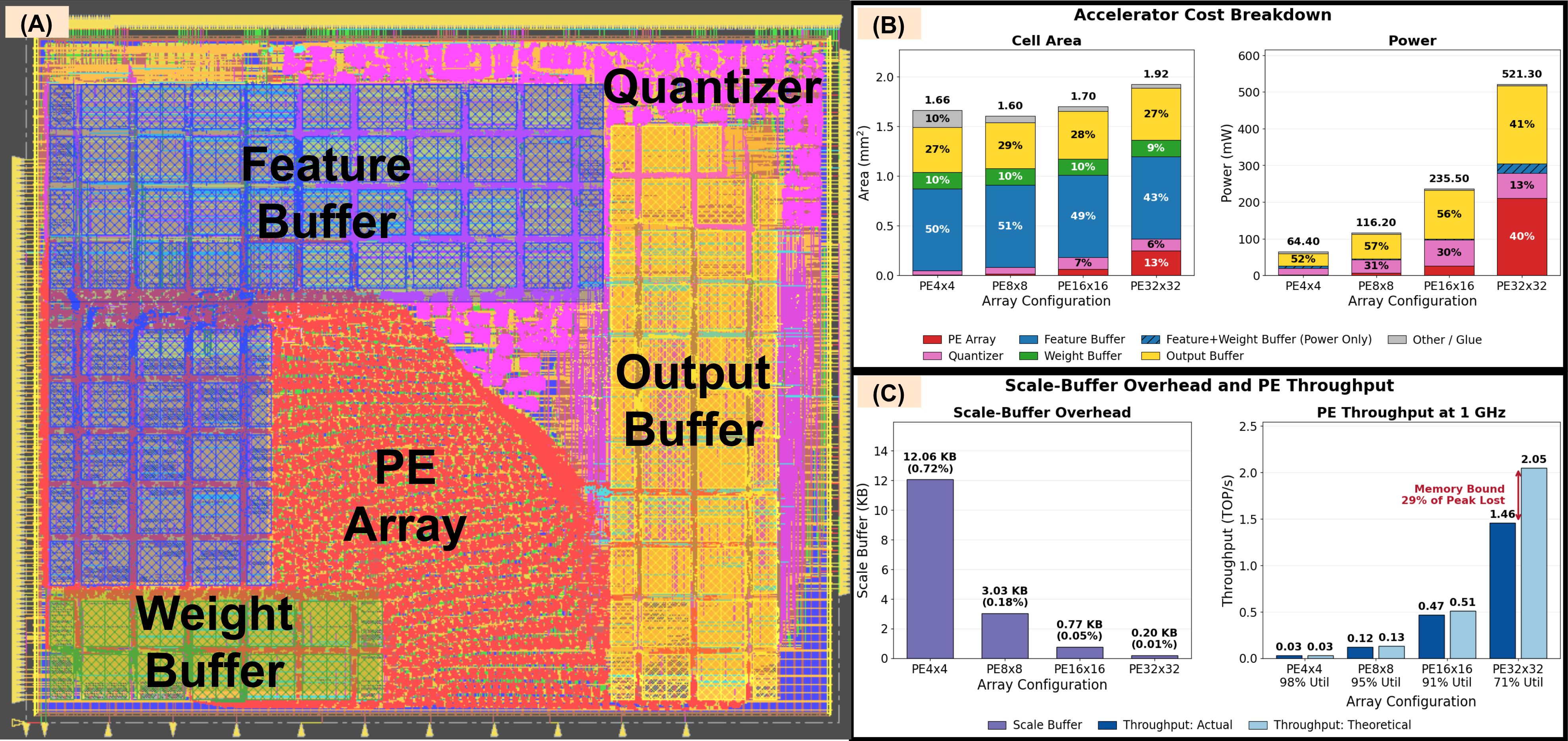}
    \caption{Hardware implementation and performance breakdown of the \sys{} Transformer Accelerator: (A) 32×32 configuration layout with color-coded components corresponding to (B) area and power breakdowns for four PE configurations, and (C) scale buffer overhead and achieved throughput relative to peak performance.}
    \label{fig:hardware_specs}
\end{figure*}

\begin{table*}[h]
\centering
\begin{tabular}{c|c|c|c|c|c|c|c|c|c}
\bottomrule
\rowcolor[HTML]{EFEFEF} 
\textbf{Work} & \textbf{Prec.} & \textbf{\begin{tabular}[c]{@{}c@{}}Tech. \\ (nm)\end{tabular}} & \textbf{\begin{tabular}[c]{@{}c@{}}Freq. \\ (MhZ)\end{tabular}} & 
\textbf{\begin{tabular}[c]{@{}c@{}}SRAM \\ (KiB)\end{tabular}} &
\textbf{\begin{tabular}[c]{@{}c@{}}Perf. \\ (TOPS)\end{tabular}}  & \textbf{\begin{tabular}[c]{@{}c@{}}Area \\ (mm$^2$)\end{tabular}} & \textbf{\begin{tabular}[c]{@{}c@{}}{Power} \\ (mW)\end{tabular}} & \textbf{\begin{tabular}[c]{@{}c@{}}Area Eff. \\ (GOPS/mm$^2$)\end{tabular}} & \textbf{\begin{tabular}[c]{@{}c@{}}Energy Eff. \\ (TOPS/W)\end{tabular}}\\ \toprule

FLTA~\cite{10595969} & A$_8$W$_8$ & 28 & 666 & 38.8 & 0.1 & 0.6 & 74 & 241 & 1.9\\

SoftEx~\cite{10971415} & BF$_{16}$ & 12 & 1120 & 256 & 0.4 & 1.21 & 581 & 355 & 1.6\\ 

FlexEdge~\cite{11561923} & A$_8$W$_8$ & 28 & 600 & 36 & 1.2 & 1.8 & 404 &  640 & 2.9\\

\cite{11075634} & INT8 & 28 & 1000 & 232 & 1.0 & 1.2 & 443 & 851 & 2.3\\

\midrule

\textbf{Ours} & \begin{tabular}[c]{@{}c@{}}A$_8$ \\ W$_{1/2/4/8}$\end{tabular} & 16 & 1000 & 1666.2 & 1.5 & 1.9 & 521 & 760 & 2.8\\ \bottomrule

\end{tabular}
\caption{Comparison of the \sys{} Transformer Accelerator with the 32x32 PE configuration against ASIC transformer works evaluated on a dual matmul workload containing a 20/40/30/10\% distribution of 1/2/4/8-bit weights.}~\vspace{-4ex}
\label{tab:asic_comparison}
\end{table*}

\subsection{Model Performance}
Focusing on the ability of \sys{} to enable intra-tensor blockwise mixed-precision quantization, Fig.~\ref{fig:deit_precision} summarizes the resulting precision distributions for a 4x4 PE configuration across the three DeiT models using a 5\% 1-bit weight target and PixelDiT-XL using 20\% 4-bit and 80\% 8-bit split. We further compare the resulting effective precision and model accuracy to existing mixed precision approaches in Tab.~\ref{tab:software_comparisons}. These results demonstrate \sys{'s} ability to assign intra-tensor precisions between 1-8-bit effective precision while maintaining competitive accuracies (max$\Delta3.3\%$) compared to coarse-grained per-tensor assignments, with a reduced overall model bitwidth. Applying our technique to PixelDit in Fig. ~\ref{fig:deit_precision} (C) shows \sys{} ability to maintain similar image generation capabilities; however, at an increased generative Fréchet Inception Distance (gFID) of 14.4 due to increased quantization noise.

\subsection{Hardware Implementation}
We implement the \sys{} Transformer Accelerator capable of executing the developed models in SystemVerilog and synthesize it under a TSMC 16nm
FinFET technology at 1~GHz. We use a 960 KiB feature, 192 KiB weight, 2 KiB bias, and 256 KiB output buffers synthesized using single-port macros and a quantizer configured with a 16-PWL segment memory and FIFOs capable of storing the required 256 columns during ViT and PixelDiT attention. Power is obtained from gate-level simulation of a
representative ViT matmul workload. We evaluate four PE array configurations,
4$\times$4 through 32$\times$32, obtained by changing the array size alone and modifying the Axi-Stream bus to standard widths of 16, 32, 128, 128 for each configuration, respectively,
while holding memory capacity and workload constant. All four configurations
meet timing.

The resulting performance breakdowns and the
placed layout of the 32$\times$32 configuration are presented in Fig.~\ref{fig:hardware_specs}. Total area ranges from 1.60
to 1.92~mm$^2$ and power from 64 to 521~mW. The design is memory-dominated.
On-chip buffers occupy 86\% of the area at 4$\times$4 and 79\% at
32$\times$32, which keeps the area nearly flat until 32$\times$32, where the
PE array first becomes significant at 13\% of area and 40\% of power. Power,
by contrast, scales with the array throughout. The quantizer, which realizes the INT32--INT8 output activation
datapath occupies only
2.6--7.1\% of total area.

Comparing the \sys{} Transformer Accelerator with comparable edge-based ASIC accelerators in Tab.~\ref{tab:asic_comparison}, \sys{} performs at a MAC throughput of 1.5 TOPS with an operating power of 521 mW. Additionally, the \sys{} Transformer Accelerator performs at a competitive area and energy efficiency of 760 GOPS/$mm^2$ and 2.8 TOPS/W. % Tab.~\ref{tab:gpu_comparison} compares the \sys{} accelerator with recent commercial GPU-based AI accelerators, demonstrating increased support for intra-tensor mixed precision at a significantly lower area and power envelope.
% \begin{table}
% \centering
% \begin{tabular}{c|c|c|c|c}
% \bottomrule
% \rowcolor[HTML]{EFEFEF}
% \textbf{Spec.} & 
% \textbf{A100} & 
% \textbf{H100} & 
% \textbf{B200} &
% \textbf{Ours} \\
% \toprule

% Precision & INT8 & INT8 & FP4 &
% A$_8$W$_{1/2/4/8}$ & 

% Tech. (nm) & 7 & 4 & 5 & 16 \\

% Perf. (TOPS) & 1248 & 3,958 & 8,875  & 1.5 \\

% Area (mm$^2$) & 826 & 814 & 
% 1628 & 1.9 \\

% Power (W) & 400  & 700 & 1430 & 0.5 \\

% Eff. (TOPS/W) & 3.1 & 5.65 & 6.2 & 2.8 \\

% \bottomrule
% \end{tabular}
% \caption{Comparison of the \sys{} accelerator with commercial GPU-based AI accelerators.}
% \label{tab:gpu_comparison}
% \end{table}

\subsection{Hardware-Software Co-Analysis}
\begin{figure}[h]
    \centering
    \includegraphics[width=\linewidth]{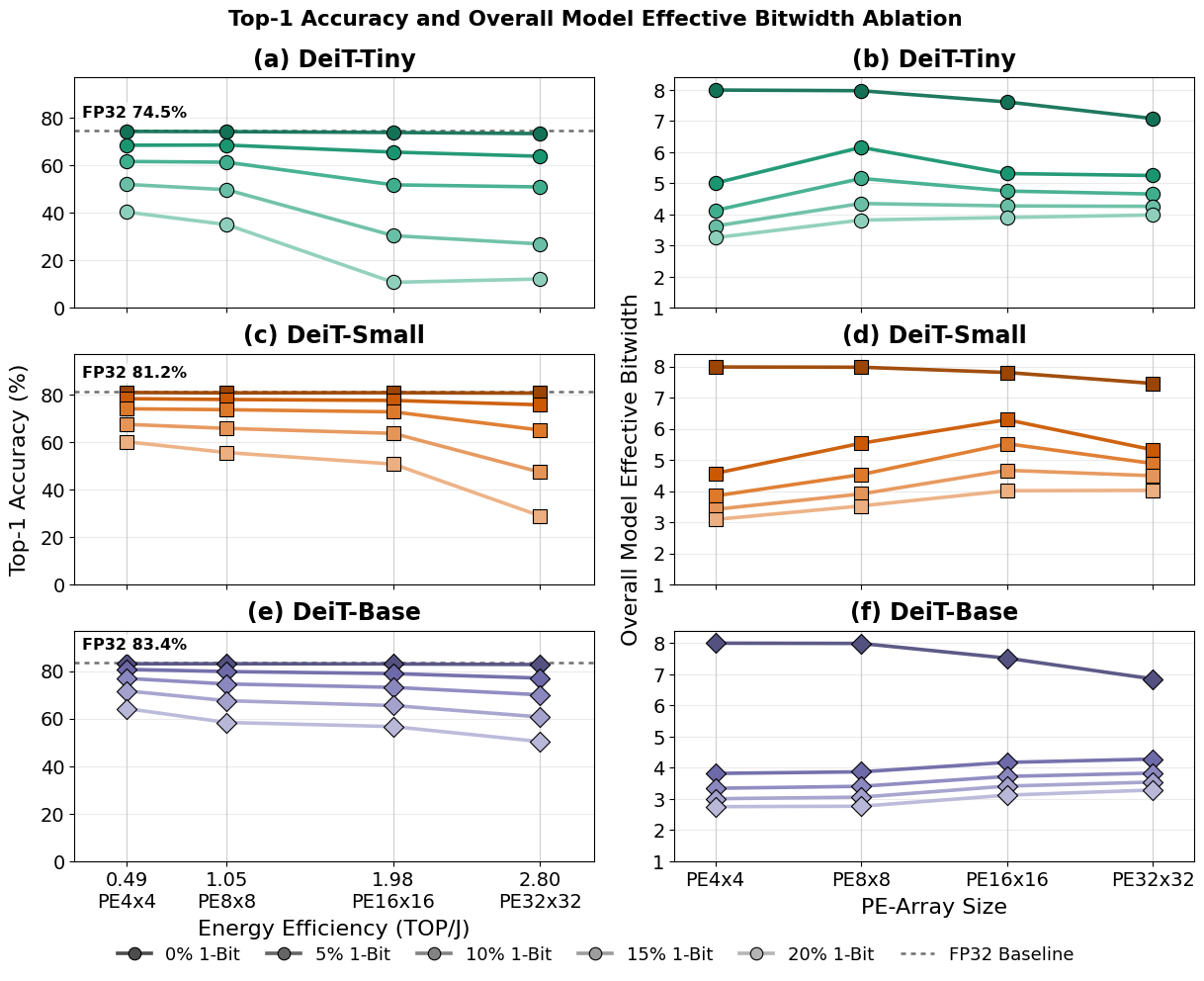}
    \caption{Top-1 accuracy vs energy efficiency (left) and overall model effective bitwidth vs PE configuration (right) for the three DeiT models tiny (a,b), small (c,d), and base (e,f). Results are shown for 5 overall 1-bit configurations of 0,5,10,15 and 20\% signified using an increasing gradient.}
    \label{fig:deit_sweep}
\end{figure}

One of the core traits of \sys{} is the correlation between the hardware systolic core configuration and performance with the software model accuracy. To demonstrate this effect, we display the top-1 accuracy for DeiT over the configuration energy efficiency within Fig.~\ref{fig:deit_sweep} for 5 target 1-bit \% configurations with the remaining percentages aligning with the CDF assignment method. Compared to the overall model performance, a 0\% 1-bit, or an assignment that is primarily dominated by 8-bit weights, leads to the highest performance for all models, approximately $<0.5\%$ within the FP model accuracy. As the 1\% distribution increases the tiny variant experiences the highest accuracy reduction, reaching approximately 40\% at a 4x4 PE configuration. On the other hand, larger models experience a smaller accuracy reduction and a reduced overall model bitwidth, with DeiT-Base reaching 60\% at an effective bitwidth of 3-bits. Additionally, as the \% of low-precision weights increases, the models experience sharper declines in accuracy and an overall increase in precision as weight-blocks experience finer weight variations. At 32×32 and 20\% 1-bit weights, the larger DeiT-Base retains substantially higher accuracy than DeiT-Tiny, indicating greater tolerance to reduced quantization granularity. Thus, larger PE arrays improve energy efficiency at the cost of model accuracy.

% Finally, with the increased parameters found within the base model, the performance drop is less pronounced comparable to an approximate 10\% top-1 in DeiT-Tiny for the 32x32 configuration and a 20\% 1-bit distribution with a $45\%$ top-1 in DeiT-Base.  This demonstrates the trade-off between hardware configuration and model performance, showing how a larger PE configuration leads to improved energy efficiency, and however, it reduces granularity within the model architecture, leading to reduced model performance. 

\begin{figure}
    \centering
    \includegraphics[width=\linewidth]{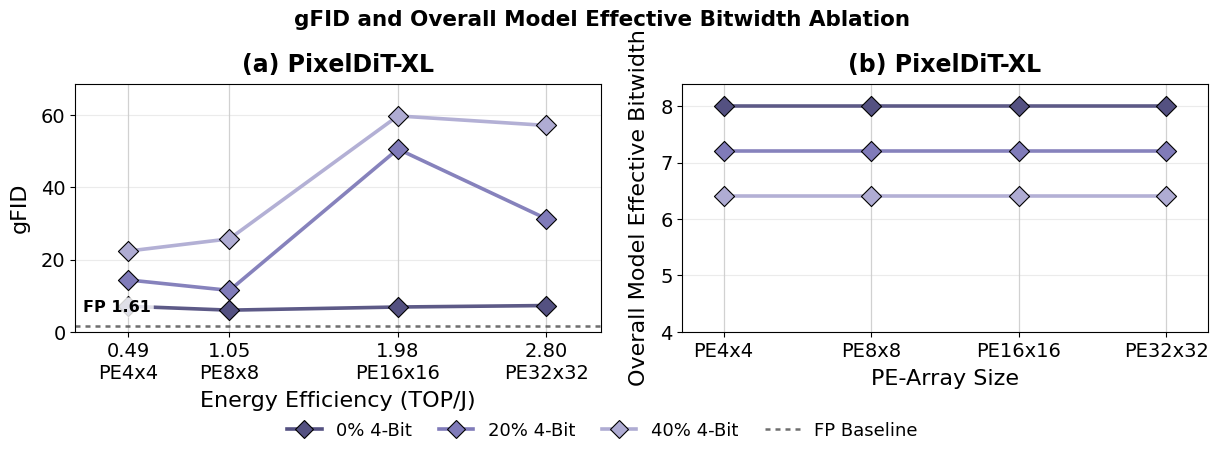}
    \caption{gFiD vs energy efficiency (a) and effective model bitwidth vs PE configuration for PixelDiT-XL (b). Results are shown for a 0,20 and 40\% 4-bit split signified by an increasing gradient.}
    \label{fig:pixeldit_sweep}
\end{figure}

We repeat a similar study for PixelDiT; however, due to the highly sensitive nature of diffusion models propagating quantization noise over tens of timesteps, we instead evaluate a higher precision split of 4- and 8-bit weights of 20\% and 80\%, as shown in Fig.~\ref{fig:pixeldit_sweep}. We emphasize the effect of increasing the PE configuration impacting overall model performance by demonstrating a gFiD increase to 50 using a 16x16 configuration, in contrast of a 14 gFiD at a 4x4 configuration using the 20\% 4-bit split.

\section{Conclusion}
This work presented \sys{}, a hardware-software co-designed framework for block-wise mixed-precision transformer inference. By jointly optimizing quantization, integer-only execution, and a configurable systolic accelerator supporting dynamic 1/2/4/8-bit computation, \sys{} enables efficient deployment of both Vision Transformers and PixelDiT image generation models. Implemented in the TSMC 16nm PDK, \sys{} achieves 1.5 TOPS, 760 GOPS/$mm^2$ area efficiency, and 2.8 TOPS/W energy efficiency, while maintaining accuracy within $3.3\%$ of state-of-the-art mixed-precision methods on DeiT at a 2-bit lower effective bitwidth, with comparable generation quality on PixelDiT. Together, these results demonstrate that fine-grained intra-tensor mixed-precision quantization can be efficiently realized through hardware-software co-design, improving the performance and flexibility of future transformer accelerators.

Future work will extend mixed-precision quantization to activations and auto-regressive transformers, and explore multi-processor integration~\cite{11261784, 9216552} to support large machine learning models. Additionally, develop approximations for advanced layers including RoPE positional embeddings, mixture-of-experts, and state-based architectures.
\section{Acknowledgments}
The authors acknowledge support from the National Science Foundation award number 2346004: Collaborative Research: NRT-Al: AI-driven next-generation semiconductor materials and devices. 

\bibliographystyle{ACM-Reference-Format}
\bibliography{eehpc}
\end{document}